\documentclass[twoside,11pt]{article}

\usepackage{blindtext}
\usepackage{amsmath}
\usepackage{dsfont}
\usepackage{tikz}
\usetikzlibrary{shapes.multipart,positioning}
\tikzset{
    block/.style = {draw, minimum height=1cm, minimum width=2cm, align=center},
    resblock/.style = {block, fill=blue!30},
    bottleneckblock/.style = {block, fill=green!30},
    arrow/.style = {->, >=latex}
}

\usepackage[preprint]{jmlr2e}

\editor{The editor}

\usepackage{lastpage}

\ShortHeadings{Human Voice Pitch Estimation}{Jeremy COCHOY}
\firstpageno{1}

\begin{document}

\title{Human Voice Pitch Estimation: A Convolutional Network with Auto-Labeled and Synthetic Data}

\author{\name Jeremy COCHOY \email jeremy.cochoy@gmail.com}

\editor{My editor}

\maketitle

\begin{abstract}%
In the domain of music and sound processing, pitch extraction plays a pivotal role. Our research presents a specialized convolutional neural network designed for pitch extraction, particularly from the human singing voice in acapella performances. Notably, our approach combines synthetic data with auto-labeled acapella sung audio, creating a robust training environment. Evaluation across datasets—comprising synthetic sounds, opera recordings, and time-stretched vowels—demonstrates its efficacy. This work paves the way for enhanced pitch extraction in both music and voice settings.

\end{abstract}


\section{Introduction}
Digital advancements have significantly impacted fields such as music and sound processing. One of the central challenges in this domain is pitch extraction from audio signals. Accurate extraction is crucial not only for transcription services but also for advanced music analysis tools and auditory perception research.

Voice pitch estimation is crucial for various applications, one of the primary being music transcription. The difficulty of this task, especially across a diverse range of instruments, has been discussed in numerous research papers. Recent studies have examined the efficacy of transformer-based models for this purpose \cite{gardner2022mt3} and variations such as the perceiver, with significant contributions highlighted in \cite{lu2023multitrack}.

A direct focus on voice pitch recognition can be found in the work of the authors in \cite{chung2023mfpam}. They employed a time-domain method combined with a meticulously crafted advanced model architecture.This led to outstanding accuracy and resilience against challenges like background noise and reverberation, albeit with a more intricate architectural design.

Traditional signal processing techniques, like Boersma's algorithm \cite{boersma1993}, have played a foundational role in pitch detection. Nevertheless, these classical techniques frequently encounter difficulties when dealing with real-world audio interferences, such as background noise and reverberation. On the other hand, recent neural network-based methods, as showcased in \cite{pin}, exhibit potential in extracting sung voice pitches and segmenting notes. Notably, these methods predominantly rely on mel spectrograms for audio-based pitch extraction.

In this study, we introduce a fully convolutional architecture designed for pitch extraction, specifically from the human singing voice. This innovative approach merges the reliability of traditional pitch detection methods with the advanced capabilities of convolutional neural networks. We employ a combination of autocorrelation and Fourier transform techniques through a Discrete Time Fourier Transform (DTFT) to encompass a broad spectrum of pitches. This inclusive approach ensures effective performance across both male and female voices, aiming to significantly enhance pitch detection accuracy in a variety of vocal ranges.

Our focus with our approach is twofold: efficiency and accuracy. We demonstrate that our streamlined architecture can achieve high accuracy levels, while being suitable for embedding in devices like smartphones for real-time pitch extraction, even in noisy environments. We anticipate that this work could inform the development of more robust note segmentation systems.

Subsequent sections detail our preprocessing methodologies, our data generation, describe our model's design, and present a thorough evaluation of its performance, concluding with a discussion of our results and their implications.

\section{Preprocessing and Data Representation}

Transcribing voice pitch from audio signifies a nuanced operation that necessitates a complex grasp of the temporal dynamics within spectral content. Various methodologies, such as those proposed in \cite{yoho} and \cite{pin}, utilize the mel spectrogram as a foundational input for processing. The introduction of the mel spectrogram is believed to capture information across various spectral domains. However, this method also leads to a loss of information, similar to what happens with pooling techniques. As a result, some valuable data may be lost. Furthermore, the spectrogram's inability to yield a precise depiction of low frequencies stems from the linear bin spacing of the Fourier transform. To mitigate this effect, and in concordance with the spectrogram, this study also incorporates autocorrelation, as detailed in \cite{boersma1993}.

\subsection{Preprocessing}

The initial stage of preprocessing involves the segmentation of the audio signal into overlapping windows, each comprising 1024 samples. This size was meticulously chosen due to its congruence with the Fast Fourier Transform (FFT) and its optimized equilibrium between computational efficiency and frequency spectrum coverage. With a sampling rate set at 44100 Hz, windows are constructed at 10 ms intervals, centered around defined focal points. The resulting 583-sample overlap between sequential windows ensures the robust preservation of signal information.

In order to amend any discontinuities manifesting at the window edges, a Hann window function is systematically applied. This procedure imposes periodicity upon the signal, priming it for the subsequent Fourier transformation.

Moreover, the computation of the volume for each respective window is conducted through the identification of the maximal absolute value within the window, furnishing an intrinsic assessment of the loudness for the specific segment. Replication of this value along the spectral dimension allow us to define a tensor with dimensions \(T \times 1024\), poised for concurrent processing with the spectral features.

The windowed segments are transformed via FFT to obtain both amplitude and phase. The amplitude, in particular, provides valuable insights into voice pitch and can be visually interpreted by trained observers.

In the context of the Short Time Fourier Transform (STFT) with overlapping windows, phase shifting occurs due to the effect of the sliding window. The phase shift is proportional to both the bin frequency and the window overlap. To counter this effect, a phase vocoder algorithm is applied to remove the phase delays between successive windows.

Given a window length of \(\ell = 1024\), hop length \(h = 441\), and a sampling frequency of 44100 Hz, the phase correction is:
\[
\phi_{w, k} = \bmod\left(\phi_{w, k} - \frac{{2 \pi w \cdot h}}{{\ell / k}}, 2\pi\right), \quad 1 \leq k < \frac{\ell}{2}
\]
where \(\phi_{w, k}\) denotes the phase of the \(k\)-th FFT bin within the \(w\)-th window.

In addition to these spectral features, the autocorrelation is computed within each window using Boersma's method, as described in \cite{boersma1993}. This technique involves normalizing the autocorrelation of the windowed signal \( R(w \cdot x; k) \) by the autocorrelation of the window \( R(w; k) \), resulting in the corrected autocorrelation:
\[
R_{\text{{corrected}}}(k) = \frac{R(w \cdot x; k)}{R(w; k)}
\]
Employing this method enhances the estimation of periodicity by taking into account the effects of the windowing process, leading to a more accurate pitch extraction.

\bigskip

\begin{figure}[ht]
    \centering
    \includegraphics[width=0.8\linewidth]{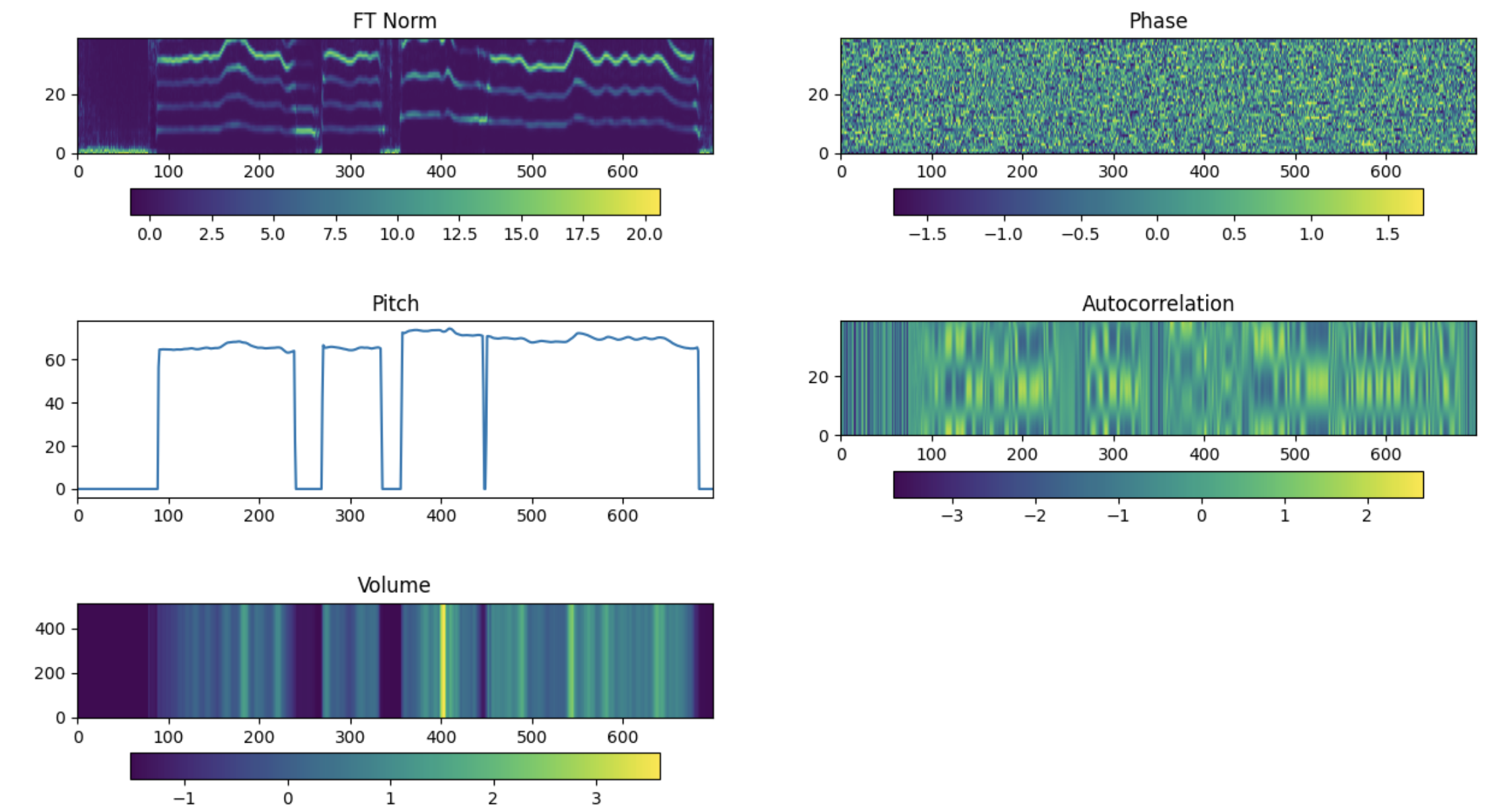}
    \caption{A visual representation of a 7s audio preprocessed input.}
    \label{fig:voiceinput}
\end{figure}

These transformations are concatenated to form a resulting tensor with a shape of \(T \times 4 \times 1024\), wherein the four channels correspond individually to amplitude, phase, autocorrelation, and volume. Due to the symmetry of the transformation, the input is truncated to a shape of \(T \times 4 \times 513\). A visual representation of this preprocessed tensor for a 7s audio input is shown in Figure \ref{fig:voiceinput}.

\subsection{Data Representation}

Music notes can be digitally represented using the MIDI standard, where each note's pitch is assigned a value between 0 and 127. This system is extended to accommodate pitches that don't align precisely with standard MIDI numbers. We represent these pitches using a 128-dimensional vector, where each element corresponds to a MIDI number and holds a value between 0 and 1. The vector effectively creates a probability distribution over the MIDI note range.

When a pitch is exactly between two MIDI numbers, like 60 and 61, the vector elements for these MIDI numbers are both assigned a value of 0.5. This represents the pitch being an equal combination of the two adjacent MIDI notes. 

To reconstruct the original pitch, denoted as \( p \), from this vector encoding, we use the formula:

\[ p = \sum_{i=0}^{127} i \cdot e(p)_i \]

In this formula, \( i \) is each MIDI number, and \( e(p)_i \) is the value in the vector for the MIDI number \( i \). The sum of the products of each MIDI number and its corresponding value in the vector gives us the original pitch value.
As an illustration, if a pitch lies between A4 (MIDI number 69) and A\#4 (MIDI number 70), the vector \( e(p) \) will have values of 0.5 at both coordinates 69 and 70, and 0 elsewhere.

This method of representation not only aligns with the physical layout of a piano keyboard but also helps the network to converge while combined with the use of Kullback-Leibler (KL) divergence loss.

The model's outputs are structured as a tensor of shape \(T \times 128\), where \(T\) represents the temporal dimension, and the second dimension corresponds to the pitch.

\section{Model Architecture}

The model architecture is inspired by the design principles of ResNet \cite{he2015deep} and WaveNet \cite{oord2016wavenet}.

The model operates on a 4-dimensional input of shape \([N, T, 4, 513]\), where \(N\) is the batch size, \(T\) is the temporal dimension, and the 4 channels correspond to amplitude of frequency, phases, autocorrelation, and volume.

The model consists of alternating stages: some are designed for dimension reduction in the spectral domain, while the others concentrate on feature extraction along the time axis, utilizing dilated convolutional layers.

\subsection{Normalization}
The input is first normalized across each instance in the batch and for each channel, specifically over the Time (T) and Frequency (F) dimensions. For each channel and instance, the mean (\(\mu\)) and standard deviation (\(\sigma\)) are computed over T and F. The normalized value \(x_{\text{norm}}\) is then computed as:
\[
\begin{aligned}
x_{\text{norm}} &= \frac{x - \mu}{\sigma + \mathds{1}_{\sigma = 0}}.
\end{aligned}
\]

\subsection{Bottleneck and Dilation Blocks}

The architecture utilizes bottleneck blocks and dilation residual blocks, which are described in the following sections.

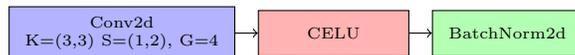
\begin{figure}[htbp]
    \centering
\begin{tikzpicture}[
    conv/.style={
        draw,
        minimum width=3cm,
        minimum height=0.6cm,
        align=center,
        font=\tiny,
        fill=blue!30,
    },
    act/.style={
        draw,
        minimum width=2cm,
        minimum height=0.6cm,
        align=center,
        font=\tiny,
        fill=red!30,
    },
    bn/.style={
        draw,
        minimum width=2cm,
        minimum height=0.6cm,
        align=center,
        font=\tiny,
        fill=green!30,
    },
]

\node[conv] (conv) {Conv2d\\ K=(3,3) S=(1,2), G=4};
\node[act, right=0.3cm of conv] (celu) {CELU};
\node[bn, right=0.3cm of celu] (bn) {BatchNorm2d};

\draw[->] (conv) -- (celu);
\draw[->] (celu) -- (bn);

\end{tikzpicture}

    \caption{Bottleneck Block}
    \label{fig:bottleneck}
\end{figure}

\paragraph{Bottleneck Blocks}
Each BottleneckBlock uses a 2D convolution layer with stride of \((1, 2)\), kernel size of \((3, 3)\), and grouped convolutions with a group parameter that is configurable (default value is 4).
This stride reduces the size of the feature tensor's last dimension, which represents 'spectral' information.
See figure \ref{fig:bottleneck}.

\begin{figure}[htbp]
    \centering
    \begin{tikzpicture}[
    conv/.style={
        draw,
        minimum width=0.75cm,
        minimum height=0.6cm,
        align=center,
        font=\tiny,
        fill=blue!30,
    },
    act/.style={
        draw,
        minimum width=0.3cm,
        minimum height=0.6cm,
        align=center,
        font=\tiny,
        fill=red!30,
    },
    bn/.style={
        draw,
        minimum width=0.5cm,
        minimum height=0.6cm,
        align=center,
        font=\tiny,
        fill=green!30,
    },
]

\node[conv] at (-1.2,0) (conv1) {Conv2d\\ K=(3,3) P=3, D=3, G=4};
\node[act, right=0.15cm of conv1] (celu1) {CELU};
\node[bn, right=0.15cm of celu1] (bn1) {BatchNorm2d};

\node[conv, right=0.15cm of bn1] (conv2) {Conv2d\\ K=(3,3) P=2, D=2, G=4};
\node[act, right=0.15cm of conv2] (celu2) {CELU};
\node[bn, right=0.15cm of celu2] (bn2) {BatchNorm2d};

\draw[->] (conv1) -- (celu1) -- (bn1) -- (conv2) -- (celu2) -- (bn2);

\draw[->] ([xshift=-1.2cm]conv1.west) -- (conv1);
\draw[->] (bn2) -- ([xshift=0.7cm]bn2.east);
\draw[->] ([xshift=-0.7cm]conv1.west) |- ([yshift=-0.5cm]bn2.south) -| ([xshift=0.3cm]bn2.east);

\end{tikzpicture}

    \caption{Dilation Block}
    \label{fig:dilation}
\end{figure}
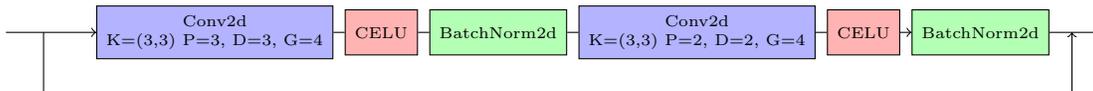

\paragraph{Dilation Residual Blocks}
DilationResBlock consists of two sequential convolution layers with dilation rates of 3 and 2. The output of the block is summed with the original input, forming a residual connection. The selection of two distinct dilation rates is intentional, aimed at facilitating effective mixing of information along the time axis. See figure \ref{fig:dilation}.

\subsection{Output Layer}

The final layers consists of a fully connected layer, followed by CELU activation and a LogSoftmax layer.

\subsection{Model Parameters}

This architecture contains approximately 1.2 million trainable parameters.

\begin{figure}[htbp]
    \centering

\begin{tikzpicture}[
    block/.style={
        draw,
        minimum width=0.75cm,
        minimum height=0.6cm,
        align=center,
        font=\tiny,
        fill=blue!30,
    },
    resblock/.style={
        draw,
        minimum width=0.75cm,
        minimum height=0.6cm,
        align=center,
        font=\tiny,
        fill=red!30,
    },
    actblock/.style={
        draw,
        minimum width=0.75cm,
        minimum height=0.6cm,
        align=center,
        font=\tiny,
    },
]

\node[block] (b1) {\begin{tabular}{c}Bottleneck\\
C=128 K=(7,7) G=1\end{tabular}};
\node[block, below=0.3cm of b1] (b2) {\begin{tabular}{c}Bottleneck\\
C=64 K=(3,3) G=2\end{tabular}};
\draw[->] (b1) -- (b2);
\node[resblock, below=0.3cm of b2] (r1) {\begin{tabular}{c}DilationResBlock\\
C=64\end{tabular}};
\draw[->] (b2) -- (r1);
\node[block, below=0.3cm of r1] (b3) {\begin{tabular}{c}Bottleneck\\
C=64 K=(3,3) G=4 \end{tabular}};
\draw[->] (r1) -- (b3);
\node[resblock, below=0.3cm of b3] (r2) {\begin{tabular}{c}DilationResBlock\\C=64\end{tabular}};
\draw[->] (b3) -- (r2);
\node[block, below=0.3cm of r2] (b4) {\begin{tabular}{c}Bottleneck\\
C=128 K=(3,3) G=4\end{tabular}};
\draw[->] (r2) -- (b4);
\node[resblock, below=0.3cm of b4] (r3) {\begin{tabular}{c}DilationResBlock\\C=128\end{tabular}};
\draw[->] (b4) -- (r3);
\node[block, below=0.3cm of r3] (b5) {\begin{tabular}{c}Bottleneck\\
C=128 K=(1,3) G=1\end{tabular}};
\draw[->] (r3) -- (b5);
\node[block, below=0.3cm of b5] (b6) {\begin{tabular}{c}Bottleneck\\
C=128 K=(1,3) G=1\end{tabular}};
\draw[->] (b5) -- (b6);
\node[actblock, below=0.3cm of b6] (conv) {\begin{tabular}{c}Conv2d\\
C=128 K=(1,5)\end{tabular}};
\draw[->] (b6) -- (conv);
\node[actblock, below=0.3cm of conv] (act2) {FC\\(C=128)};
\draw[->] (conv) -- (act2);

\end{tikzpicture}

    \caption{Neural Network Architecture}
    \label{fig:architecture}
\end{figure}
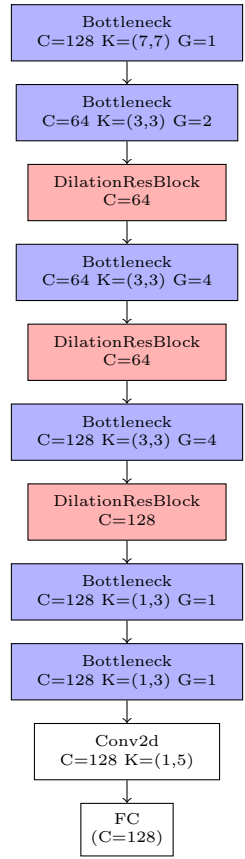

\section{Dataset and Training}

\subsection{Datasets}
A significant part of our research centers on the data employed to train our model. Instead of depending on manually labeled data, we utilize a combination of synthetic data, where audio is generated based on a label, and ground truth audio from acapella singers. These are automatically labeled using a C implementation of Boersma's algorithm. Although the labels generated in this second process are somewhat noisy, neural networks are well-suited for generalizing from such data.

\paragraph{Synthesizer Samples Generation}

The initial dataset in our study is entirely synthetic, resembling sounds typical of sequencers and synthesizers. Our methodology involves generating audio waves through additive or subtractive synthesis, based on sequences of notes with specified amplitudes. The procedure commences with the uniform sampling of paris of pitches and note durations to create a musical sequence, akin to a music score.
Subsequently, each note in this sequence is converted into audio by generating a sound wave randomly salected (uniform distribution) among one of the following waveforms: sine, triangle, square, or sawtooth.

Following the generation phase, there is a 30\% chance of applying a randomly chosen low-pass filter, with the cutoff frequency uniformly sampled between 1000 and 20000 Hz. This step emulates typical synthesizer filters. Additionally, we introduce Gaussian noise with an amplitude of 10\% to the synthesized audio.

While our process could be further enhanced by incorporating various effects or noises, such as delay, echo, reverberation, or ambient sounds from environments like restaurants or pubs, these aspects were not explored in depth within the current scope of our research. However, we hypothesize that such enhancements could bolster the model's robustness and potentially synergize with different modeling approaches or time-only methods, given their independence from the specific neural network architecture employed.

\paragraph{Chinese and Western Opera Dataset}
The second dataset features predominantly Chinese Opera recordings, supplemented by some samples from Western Opera, as referenced in \cite{opera}. The audio files are divided into segments lasting 7 seconds. Segments that fall short of this length are padded with silence. The pitch labels for these segments are computed using our C implementation of Boersma's algorithm. Despite the presence of imperfections and false negatives within the dataset, the neural network exhibits the ability to generalize, thus overcoming the noise present in the labels.

\paragraph{Time-Stretched Synthetic Vowel Dataset}
The third dataset is synthetically constructed by "stretching" the audio signal in the time domain of samples from the North Texas vowels database, as referred to in \cite{texas}. This manipulation inevitably alters the original pitch.

The method for generating this dataset is akin to the procedure used for the initial synthesizer dataset. The process begins by creating non-overlapping notes of various lengths and positions, without assigning any pitch. Following this, a sample is selected from a vowel database and is elongated to match the note's duration. Although attempts are made to eliminate silences at the beginning and end of these samples, some noise susceptibility remains. Consequently, the audio may not align perfectly with the intended note duration due to residual silence in the original recordings.

In the final step, Boersma's algorithm is utilized for automatic pitch labeling. This approach accounts for the pitch alterations caused by the stretching of the samples. An alternative method, not explored in this study, involves using an auto-tune algorithm to adjust the pitch to specific values.

\medskip

In summary, the datasets generated for this study include approximately 5.8 hours of synthetic notes, 2.8 hours of synthetic voice audio, and 5.7 hours of opera recordings. All the data in this study was generated without relying on human labeling.

\subsection{Model Training}
The model training process is carried out using the three distinct datasets previously described. Each dataset is partitioned into three subsets: 50\% allocated for training, 25\% designated for validation, and the remaining 25\% reserved for testing.

We employ the Kullback-Leibler (KL) divergence as the loss function to measure the dissimilarity between the log-softmax output of our model and the one-hot encoded pitch label. This loss is computed only for instances where the pitch labels do not correspond to silence, i.e., for non-zero pitch labels. The KL divergence in this context is defined as:
\begin{equation}
D_{\text{KL}}(y \parallel \hat{y}) = \sum_{i} \mathds{1}_{(y_i \neq 0)} y_i \log \frac{y_i}{\hat{y}_i}
\end{equation}
where \( y \) represents the true one-hot encoded pitch labels, and \( \hat{y} \) represents the predicted log probabilities output by the model.

The model's training is carried out using the Adam optimizer, with a mini-batch size of 16 and a learning rate of \(10^{-3}\). We apply the standard initialization method provided by PyTorch. It is worth noting that, in our experimentation, we did not conduct any hyperparameter optimization.

\section{Experimental Analysis}

\subsection{Model Selection Criteria}
During our experimentation, it was observed that the model begins to overfit post approximately 20 epochs. Hence, the optimal model was chosen based on the loss computed on the validation dataset prior to any subsequent evaluations.

\subsection{Evaluation Metrics}
To quantify the model's performance, we adopted multiple metrics. Specifically, we computed the mean, median, 25th, 75th, and 99th percentiles of the absolute discrepancies between the model's pitch predictions and the true labels, across three distinct validation datasets. This was only carried out in instances where the true labels were available.

Furthermore, we assessed the model's accuracy using a 50 cents pitch precision, which is equivalent to a tolerance of a half semitone, for each validation dataset. These results are collated in Table \ref{fig:non_delayed_metrics}.

A supplementary analysis was conducted using a temporal tolerance of 30ms. This involved comparing the model's predictions with three versions of labels: the original, one shifted by +10ms, and another by -10ms. For each prediction, the nearest label from these three was chosen for comparison, restricting to timestamps where labels were non-zero. The outcomes of this evaluation are detailed in Table \ref{fig:delayed_30ms_metrics}.

All accuracy measures are represented in percentage values, whereas pitch discrepancies are expressed in cents.

\begin{table}[h]
\centering
\begin{tabular}{|l|c|c|c|c|c|c|c|}
\hline
Dataset & Acc. & Err. Mean & Err. 25th & Err. Median & Err. 75th & Err. 99th \\
\hline
Synth   & 99.47 & 15.96 & 4.38 & 9.26 & 15.85 & 40.61 \\
Voice   & \textbf{94.03} & 84.64 & 3.31 & 7.50 & 15.39 & 3005.10 \\
Sampled & 88.22 & 188.33 & 2.90 & 6.57 & 14.13 & 4329.29 \\
\hline
\end{tabular}
\caption{Performance comparison across datasets using non-delayed metrics.}
\label{fig:non_delayed_metrics}
\end{table}

\begin{table}[h]
\centering
\begin{tabular}{|l|c|c|c|c|c|c|c|}
\hline
Dataset & Acc. & Err. Mean & Err. 25th & Err. Median & Err. 75th & Err. 99th \\
\hline
Synth  & 99.56 & 13.04 & 4.36 & 9.24 & 15.82 & 39.86 \\
Voice  & \textbf{97.75} & 19.44 & 1.38 & 3.37 & 7.04 & 438.00 \\
Sampled & 92.90 & 57.47 & 1.28 & 3.51 & 8.22 & 1206.20 \\
\hline
\end{tabular}
\caption{Performance comparison across datasets using 30ms delayed metrics.}
\label{fig:delayed_30ms_metrics}
\end{table}

\subsection{Ablation Study}
\begin{figure}[h]
\centering
\includegraphics[width=0.4\linewidth]{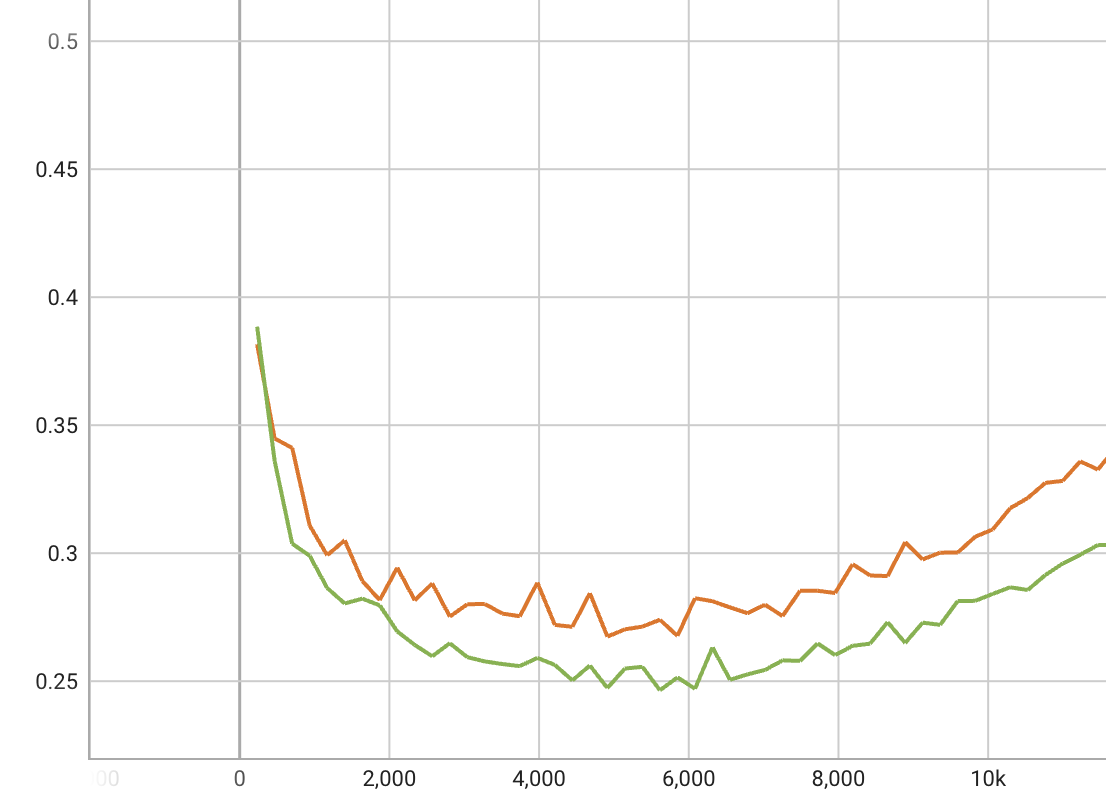}
\caption{Comparison of validation loss: Ablation vs. Original Configuration.}
\label{fig:validation_loss_comparison}
\end{figure}

To validate the hypothesis that autocorrelation enhances the model's effectiveness, we performed an identical training regimen, nullifying the autocorrelation channel for all input datasets, encompassing both training and test samples. A discernible surge in validation loss was observed, as illustrated in Figure \ref{fig:validation_loss_comparison}.

Subsequent evaluations using our established metrics were also conducted and are presented in Tables \ref{fig:ablation2_non_delayed_metrics} and \ref{fig:ablation2_delayed_metrics}.

\begin{table}[h]
\centering
\begin{tabular}{|l|c|c|c|c|c|c|}
\hline
Dataset & Acc. & Err. Mean & Err. 25th & Err. Median & Err. 75th & Err. 99th \\
\hline
Synth   & 99.08 & 19.69 & 4.74 & 10.08 & 17.53 & 48.02 \\
Voice   & 93.56 & 86.07 & 4.44 & 9.77 & 18.79 & 3038.81 \\
Sampled & 88.24 & 200.98 & 3.47 & 7.80 & 16.29 & 4395.98 \\
\hline
\end{tabular}
\caption{Datasets comparison post-ablation using non-delayed metrics.}
\label{fig:ablation2_non_delayed_metrics}
\end{table}

\begin{table}[h]
\centering
\begin{tabular}{|l|c|c|c|c|c|c|}
\hline
Dataset & Acc. & Err. Mean & Err. 25th & Err. Median & Err. 75th & Err. 99th \\
\hline
Synth   & 99.22 & 15.41 & 4.71 & 10.05 & 17.48 & 45.80 \\
Voice   & 97.75 & 18.59 & 1.82 & 4.46 & 9.23 & 310.88 \\
Sampled & 93.92 & 55.28 & 1.54 & 4.28 & 9.72 & 1206.76 \\
\hline
\end{tabular}
\caption{Datasets comparison post-ablation using delayed metrics.}
\label{fig:ablation2_delayed_metrics}
\end{table}

\section{Conclusion}

In our research, we presented a fully convolutional neural network adept at delivering precise pitch estimations of human voice, harnessing the combined capabilities of autocorrelation and spectrogram. The inherent simplicity and efficiency of this architecture are noteworthy.

The introduction of synthetic data has significantly amplified the model's performance, instilling it with greater resilience against noisy backgrounds. This adaptability showcases the potential of the framework to effectively process acapella recordings even in challenging environments like cafes or restaurants.

Future enhancements to the model architecture can take the form of a segmentation head, purposed for the accurate detection of note onsets and offsets as well as distinguishing "silences" — sections of audio recordings rife with irrelevant noise and information unrelated to our primary task. 

On the dataset front, advancements could be realized by refining the vowel stretching note generation via a vocoder. This would not only allow for better pitch control but also facilitate accurate labeling for pitch, onset, and offset. Furthermore, integrating a consonant at the onset presents an exciting avenue for exploration.

\vskip 0.2in
\bibliography{pitchnet}

\end{document}